\documentclass[prl,twocolumn,showpacs]{revtex4}
\usepackage[xdvi]{graphicx}%
\usepackage{dcolumn}
\usepackage{epsfig}

\newcommand{\eqref}[1]{(\ref{#1})}

\begin{document}
\title{Time-symmetric fluctuations in nonequilibrium systems}
\author{Christian Maes${}^{1}$ and Maarten H. van Wieren${}^{2}$}
\affiliation {${}^1$Instituut voor Theoretische Fysica, K.U.Leuven
\\
${}^2$EURANDOM, Eindhoven}

\date{\today}


\begin{abstract}
For nonequilibrium steady states, we identify observables whose
fluctuations satisfy a general symmetry and for which a new
reciprocity relation can be shown. Unlike the situation in
recently discussed fluctuation theorems, these observables are
time-reversal symmetric. That is essential for exploiting the
fluctuation symmetry beyond linear response theory. Besides
time-reversal, a crucial role is played by the reversal of the
driving fields, that further resolves the space-time action.  In
particular, the time-symmetric part in the space-time action
determines second order effects of the nonequilibrium driving.
\end{abstract}


\maketitle


Nonequilibrium statistical mechanics is still very much under
construction.  Its very rich phenomenology reaching from some of the great
unsolved problems of classical physics such as turbulence, to the
interdisciplinary fields of biophysics and complex systems, has not yet
found a sufficiently general and powerful theoretical foundation. Compared
with equilibrium statistical mechanics, we lack a global fluctuation
theory of far from equilibrium systems, we do not have a controlled
perturbation theory and even many conceptual debates have not found their
generally accepted and logical conclusions. In contrast, the Gibbs
formalism of equilibrium statistical mechanics yields general identities
and inequalities connecting thermodynamic quantities and correlation
functions. They relate the response of a system to an external action with
the fluctuations in the system of the corresponding variable. That is
partially inherited by irreversible thermodynamics, as in the
fluctuation-dissipation theorem, \cite{deGM}, that works close to
equilibrium but is mostly limited to the so called linear response regime;
away from equilibrium no such general relations exist.

In recent years however \cite{ev,GC,Jarz,poincare,PT,ber1,bod1} a broader
range of thermodynamic and fluctuation relations have become available
also further away from equilibrium and there has been a general sentiment
that these are important steps in the construction of nonequilibrium
statistical mechanics. In what follows we limit ourselves to the steady
state. A fluctuation symmetry for a dissipation function $S$ has been
proposed and studied in various contexts. Schematically, that fluctuation
relation goes somewhat like
\begin{equation}
\mbox{Prob}[S=\sigma]=\mbox{Prob}[S=-\sigma]\,\exp\sigma\label{FT}
\end{equation}
where the probability distribution is in the steady state and $S$ denotes
in general a path-dependent variable whose average can often be
interpreted as  the total change (in dimensionless units) of entropy over
a long time interval $[0,\tau]$. Depending on the particular realization,
the function $S$ is better called entropy current, dissipated work or
entropy production. Subject to the scale of description
$S=S_\varphi(\omega)$ is a function of the path or history $\omega$ and of
the driving fields $\varphi$. We will call $S$ the dissipation function.

Symmetries such as \eqref{FT} first appeared in \cite{ev,GC} in the
context of thermostated and smooth dynamical systems where the phase space
contraction is identified with entropy production. A slightly stronger
statement is that the steady state averages satisfy
\begin{equation}\label{ts}
 \langle f\theta\rangle=\langle f\, e^{-S}\rangle
\end{equation}
where $f=f(\omega)$ is a function of the history
$\omega=(\omega_{t})_{0}^{\tau}$ of the system over the time interval
$[0,\tau]$ and $\theta$ is the time-reversal,
$\theta\omega=(\omega_{\tau-t})_{0}^{\tau}$, so that $ f\theta(\omega)=
f(\theta\omega)$.

While that symmetry \eqref{ts} is very generally valid and while the
symmetry-identity \eqref{FT} is non-perturbative (i.e., not restricted to
close to equilibrium situations), so far its further theoretical
consequences have only been made visible in the derivation of linear
response relations such as those of Green and Kubo, see e.g.\
\cite{Gr,maes}. The bad news is that while relations such as \eqref{FT} or
\eqref{ts} are very general, their consequences for expansions around
equilibrium are necessarily limited to first order, the linear regime, as
we show next.

{\bf Expansion of the fluctuation symmetry for the dissipation function.}
As a central assumption, here as in irreversible thermodynamics
\cite{deGM} we take the dissipation function $S$ as given by
\begin{equation}\label{eq:S_in_phi_and_J}
S =  S_\varphi(\omega) = \sum_{\alpha} \varphi_\alpha
J_\alpha(\omega)=\varphi\cdot J
\end{equation}
where the $J_\alpha$'s are the currents describing the displacement of a
quantity of type $\alpha$ (matter, charge, energy,...); the
$\varphi_\alpha$'s indicate the various accompanying (driving) fields or
affinities ($\varphi= 0$ is set to correspond to equilibrium). This
assumption implies that we ignore temporal boundary terms of the form
$U(\omega_\tau) - U(\omega_0)$, as they usually play no further role in
the fluctuation theory for $\tau\uparrow +\infty$, see however
\cite{cohenvanZon}. We write $\langle\cdot\rangle_\varphi$ for the steady
state expectation and $\langle\cdot\rangle_0=
\langle\cdot\rangle_{\mbox{eq}}$ stands for the equilibrium expectation.
No systematic expansion around equilibrium is at all possible from the
relations \eqref{FT} or \eqref{ts}. That can be seen as follows.

If $f^{+}$ is time-symmetric, $f^{+}\theta=f^{+}$, then \eqref{ts} gives
to linear order in $\varphi$,
\begin{equation}
 \langle f^{+}\rangle_\varphi=\langle f^{+}\, e^{-S}\rangle_\varphi \Rightarrow
 \langle f^{+}\rangle_\varphi=
 \langle f^{+}\rangle_\varphi-\langle f^{+}\, \varphi\cdot J\rangle_0
 \label{tsS}
\end{equation}
 and indeed, the product $Jf^{+}$ is antisymmetric
under time-reversal, hence vanishes under equilibrium expectation:
\eqref{tsS} results in the identity $0=0$ to first order in $\varphi$ and
\eqref{ts} gives no first order term around equilibrium for time-symmetric
observables.  If $f^{-}$ is antisymmetric, $f^{-}\theta=-f^{-}$, then the
second order term of $\langle f^-\rangle_\varphi$ requires information
about the first order terms $\langle f^{-}J_\alpha\rangle_\varphi$ but
that information is lacking because of the arguments above applied to the
time-symmetric observable $f^{+}=f^{-}J_\alpha$.

{\bf Lagrangian set-up.} From the above remarks, it is clear what is
missing. We need information about the generation of time-symmetric
observables. That is made most visible in a Lagrangian set-up.  It amounts
to the realization of the steady state space-time distribution as a Gibbs
distribution.  That can be done under a wide variety of contexts, see
\cite{kubo,ber1,bod1,poincare,maes,GC,om}, but most easily and explicitly
for stochastic dynamics.

Quite generally we can construct the probability distribution
$P_{\varphi}$ on space-time configurations $\omega$ for a nonequilibrium
steady state parameterized by some vector $\varphi=(\varphi_\alpha)$, in
terms of the corresponding equilibrium distribution $P_{0}$:
\begin{equation}\label{action}
P_{\varphi}(\omega)=\exp[-\mathcal{L}_{\varphi}(\omega)]\,
P_{0}(\omega)\label{lan}
\end{equation}
denoting the Lagrangian action by $\mathcal{L}_\varphi$ to avoid confusion
with the dissipation function $S$. One cannot go too lightly over the
choice of the equilibrium distribution $P_{0}$ but we momentarily choose
to ignore these issues.

Relation \eqref{ts} teaches that the time-antisymmetric part is given by
the dissipation function:
\begin{equation}\label{lananti}
 S_\varphi(\omega) =\mathcal{L}_{\varphi}(\theta\omega) - \mathcal{L}_{\varphi}(\omega)
\end{equation}
Indeed, with \eqref{lananti}, \eqref{ts} is equivalent with
\begin{equation}\label{rd}
 P_{\varphi}(\omega)=e^{S_{\varphi}(\omega)}P_{\varphi}(\theta\omega)
\end{equation}
where we used that in equilibrium, $P_{0}\theta=P_{0}$; that relation
implies \eqref{FT}. In other words, the source of time-reversal breaking
is (essentially) given by the dissipation function. That has been
confirmed by many physically motivated examples and models and was also
derived in much greater generality \cite{mn,poincare}. It was also argued
there how \eqref{rd} can be interpreted as a unification of existing
fluctuation theorems, ranging from the Jarzynski equality \cite{Jarz} to
the Gallavotti-Cohen theorem for chaotic dynamical systems \cite{GC}.

To go beyond linear order in the nonequilibrium response-functions, it
appears that we must obtain extra information about the time-symmetric
part in $\mathcal{L}_{\varphi}$. It is important to understand here that
there is no reason to think that the nonequilibrium driving would not
generate an extra term in $\mathcal{L}_\varphi$ which is symmetric under
time-reversal. Adding a nonequilibrium driving changes the time-symmetric
part in the space-time action, which is the reason why the response
functions cannot be generated by the expression for the dissipation
function (or entropy production) alone.

{\bf Time-symmetric part.} For characterizing nonequilibrium one has to
realize that time-symmetry breaking is not spontaneous and is itself very
much linked with, if not caused by, breaking of spatial or even internal
symmetries. A simple case is steady heat conduction made possible by the
spatial arrangement of different heat baths in contact with the system.
Therefore, to resolve further the nonequilibrium state, one can exploit
also directly the (anti-)symmetries associated to the driving fields. It
is straightforward to introduce the notion of field-reversal as
 $\varphi_\alpha\rightarrow -\varphi_\alpha$,
which could e.g. mean to exchange the two heat reservoirs in a
one-dimensional heat conduction problem or to reverse the direction of an
externally given driving field.  We utilize field reversal as our basic
transformation on the driving fields and with \eqref{eq:S_in_phi_and_J}
\begin{equation}\label{eq:S_pi}
 S_{-\varphi}= -S_{\varphi} \mbox{, and hence }\;
 S_\varphi(\theta\omega) = S_{-\varphi}(\omega)
\end{equation}
This is different for the time-symmetric part in the Lagrangian $A_\varphi
\equiv \mathcal{L}_{\varphi} + \mathcal{L}_{\varphi}\theta$, for which in
general $A_\varphi \neq \pm A_{-\varphi}$. For
$Y_\varphi\equiv\mathcal{L}_{-\varphi} - \mathcal{L}_{\varphi}$  we have
the identity $A_{-\varphi} - A_\varphi = Y_\varphi\theta + Y_\varphi$ and
we may derive that $A_\varphi=A_{-\varphi}$ iff $Y_\varphi=S_\varphi$ iff
$ \mathcal{L}_{-\varphi} = \mathcal{L}_{\varphi}\theta$.

The reason why the time-symmetric term $A_\varphi$  is unseen in the
linear response to currents is because, to linear order, field-reversal
can be replaced with time-reversal: for small $\varphi$,
\[
\langle J \rangle_{-\varphi} \simeq -\langle J\rangle_{\varphi} =
\langle J\theta\rangle_{\varphi}
\]
This is not generally true beyond order $\varphi$, which is for example
all-important for the construction of so called rectifiers, see for
instance \cite{rec}. Hence, while for small driving (small $\varphi$,
close to equilibrium), field-reversal can be implemented by time-reversal,
the time-symmetric $A_\varphi$ is expected to be important further away
from equilibrium. Consider for example the flow of a viscous fluid through
a tube under influence of a pressure difference $\varphi=\Delta p$. For
small $\varphi$ one observes a laminar flow and hence, according to
Poiseuille's equation the flow rate is proportional to the pressure drop
and one cannot distinguish between reversal of time and reversal of the
field or pressure difference. At higher pressure differences however, or
above a critical velocity, the laminar pattern breaks up and the flow
becomes turbulent. Then time-reversal of the flow pattern definitely
differs from the typical pattern obtained via field-reversal, and hence
$\mathcal{L}_\varphi$ must have a time-symmetric part. Ultimately,
$A_\varphi\neq A_{-\varphi}$ ($\Leftrightarrow \mathcal{L}_{-\varphi} \neq
\mathcal{L}_\varphi\theta$) is intimately related to nonlinear response.
The \emph{field}-antisymmetric part $A_{-\varphi}- A_{\varphi} =
Y_\varphi\theta + Y_\varphi$ rather than the \emph{time}-antisymmetric
part (the dissipation function $S_\varphi$) is also the pivotal quantity
to observe, when one attempts higher order expansions.

{\bf New fluctuation symmetry.} Remember that $Y_\varphi =
\mathcal{L}_{-\varphi} - \mathcal{L}_\varphi$. Applying field
reversal we have, by construction,
\begin{equation}\label{phisfs}
\langle f\rangle_{-\varphi} = \langle f
e^{-Y_\varphi}\rangle_\varphi
\end{equation}
Combining it with time-reversal,
\begin{equation}\label{phifs}
\langle f\theta \rangle_{-\varphi} = \langle f e^{-\frac
1{2}(Y_\varphi\theta + Y_\varphi)}\rangle_\varphi
\end{equation}
where we have used that $S_\varphi = (Y_\varphi - Y_\varphi\theta)/2$. As
a consequence, for $R_\varphi\equiv(Y_\varphi + Y_\varphi\theta)/2$ we
have Prob$_\varphi[R_\varphi=r] = $Prob$_{-\varphi}[R_\varphi=r]\,e^{r}$.
It often happens that the field reversal can be implemented as a map on
the histories.  In that case, there is an involution $\Gamma$ on paths
$\omega$ so that $\mathcal{L}_{-\varphi} = \mathcal{L}_\varphi \Gamma$.
The transformation $\Gamma$ could for example simply spatially mirror all
the internal degrees of freedom, cf. the examples below. Then, from
\eqref{phifs} and similar to \eqref{ts},
\begin{equation}\label{phifs2}
\langle f\theta\Gamma \rangle_{\varphi} = \langle f e^{-\frac
1{2}(Y_\varphi\theta + Y_\varphi)}\rangle_\varphi
\end{equation}
As $R=R_\varphi$ satisfies $R\theta\Gamma=-R$, we then have the
fluctuation symmetry \eqref{FT} not only for the dissipation function $S=
(Y_\varphi - Y_\varphi\theta)/2$ but we now also get it for
$R\equiv(Y_\varphi + Y_\varphi\theta)/2 = (A_{-\varphi} - A_\varphi)/2$:
\begin{equation}\label{FSS}
\mbox{Prob}[R=r]=\mbox{Prob}[R=-r]\,e^{r} \label{FTS}\end{equation} where
the probability distribution is in the steady state and the fluctuation
observable $R$ denotes the variable time-symmetric and
field-reversal-antisymmetric part in the Lagrangian action over a long
time-interval $[0,\tau]$. This fluctuation symmetry is again generally
valid, non perturbatively and away from equilibrium. This time however, it
has implications for the response of time-symmetric observables. Indeed,
from \eqref{phisfs} or \eqref{phifs}, when $f^+ = f^+\theta$,
\[
\frac 1{2\varphi}\big(\langle f^+\rangle_\varphi - \langle
f^+\rangle_{-\varphi}\big) = \frac 1{2\varphi}\langle
f^+(1-e^{-R})\rangle_\varphi
\]
\begin{equation}\label{1st}
\frac{\partial}{\partial \varphi_\alpha}\langle f^+\rangle_{\varphi=0} =
\frac 1{2}\langle f^+\,\frac{\partial}{\partial \varphi_\alpha}R_\varphi
\rangle_{\varphi=0}
\end{equation}
As a result, we have an Onsager reciprocity for the observables
\begin{equation}\label{vons}
V_\alpha\equiv \frac{\partial}{\partial
\varphi_\alpha}R_\varphi|_{\varphi=0}
\end{equation}
in the sense that to first order $\langle V_\alpha \rangle_\varphi =
\langle V_\alpha \rangle_0 + \sum_\gamma M_{\alpha\gamma} \varphi_\gamma$
with symmetric linear response coefficients
$M_{\alpha\gamma}=M_{\gamma\alpha}$; this can be seen by taking
$f^+=V_\gamma$ in \eqref{1st}. The application of \eqref{1st} obviously
leads to a higher order expansion of the currents $J_\alpha$ around
equilibrium when taking $f^+=J_\alpha J_{\gamma}$ in \eqref{1st}.

{\bf Examples.} We consider here a classical model of heat conduction,
\cite{eck1,mnv}. At each site $i=1,\ldots,N$ there is an oscillator
characterized by a scalar position $q_i$ and momentum $p_i$.  The dynamics
is Hamiltonian except at the boundary $\{1,N\}$ where the interaction with
the reservoirs has the form of Langevin forces as expressed by the It\^o
stochastic differential equations
\begin{eqnarray}\label{eqmo}
dq_{i}&=&p_{i}dt, \quad i=1,\ldots,N\nonumber\\
dp_{i}&=&-\frac{\partial U}{\partial q_{i}}(q)dt, \quad i=2,\ldots,N-1\nonumber\\
dp_{i}&=&-\frac{\partial U}{\partial q_{i}}(q)dt-\gamma
p_{i}+\sqrt{\frac{2\gamma}{\beta_i}}dW_i(t), \quad i=1,N\nonumber
\end{eqnarray}
The $\beta_1,\beta_N$ are  the inverse temperatures of the heat baths
coupled to the boundary sites $i=1,N$; $dW_1(t),dW_N(t)$ are mutually
independent standard white noise. Appropriate growth and locality
conditions on the potential $U$  (which need not be homogeneous) allow the
existence of a corresponding smooth Markov diffusion process with a unique
stationary distribution. When $\beta_1=\beta_N=\beta$, the Gibbs measure
$\sim\exp[-\beta H], H = \sum p_i^2/2 + U$ is time-reversible for the
process with kinematical time-reversal $\pi$ given by the involution $\pi
f(q,p)=f(q,-p)$.  Let $\omega=((q(t),p(t)),t\in [0,\tau])$ denote the
evolution of the system  in the period $[0,\tau]$. The natural definition
of time-reversal $\theta$ is thus $(\theta\omega)_t=\pi(\omega_{\tau-t})=
(q(\tau-t),-p(\tau-t))$.  For determining the action in \eqref{action} we
take for $P_0$ the equilibrium process where the inverse temperatures of
both heat reservoirs are equal to $\beta=(\beta_1 +\beta_N)/2$. There is
one driving field $\varphi = \beta_1-\beta_N$. To compute the
antisymmetric part under time-reversal, see \eqref{lananti}, in \cite{mnv}
standard stochastic calculus yields $S_\varphi = (\beta_1-\beta_N) J$
where $J= J(\omega)$ is the (path-dependent) heat dissipated in the
reservoir at site $i=1$ in time $\tau$. The steady state average of
$S_\varphi$ is the heat dissipation and is easily seen to be non-negative
by integrating \eqref{rd} over all $\omega$ and by using the Jensen
inequality, see further in \cite{eck1,luc1,mnv}. But we can also compute
the symmetric part under time-reversal; it equals
\[
A_\varphi(\omega) =
\frac{\beta_1^2-\beta^2}{\beta}\int_0^\tau\,p_1^2(t)dt +
\frac{\beta_N^2-\beta^2}{\beta}\int_0^\tau\,p_N^2(t)dt
\]
which involves the time-integral of the (path-dependent) kinetic
temperatures at the ends of the chain.  Field reversal corresponds to
exchanging $\beta_1$ with $\beta_N$ and hence
\[
R= (\beta_N-\beta_1)\,\big[\int_0^\tau\,p_1^2(t)dt
-\int_0^\tau\,p_N^2(t)dt\big]
\]
satisfies the fluctuation symmetry \eqref{FSS}.  Furthermore, for
\eqref{vons}, we find
\[
V = \int_0^\tau\,p_N^2(t)dt -\int_0^\tau\,p_1^2(t)dt
\]
from which we get that the left/right difference of time-integrated
kinetic temperatures satisfies a Green-Kubo type relation \eqref{1st}, but
now for the time-symmetric observable $V = \int (p_N^2 - p_1^2)dt $,
\[
\frac{\partial}{\partial \varphi}\langle V\rangle_{\varphi=0} =
\frac 1{2}\langle V^2\rangle_{\varphi=0}
 \]

A second class of examples is given by the more general Langevin-type
equation
\[
dx(t)= -F(x(t))\,dt + \sqrt{2}\,dW(t)
\]
The drift is given by the force $-F$, that, in nonequilibrium situations,
cannot be derived from a potential. To parameterize the driving, we write
$F=\nabla U - \varphi G$ where U is a potential function and $G$ is an
external force that possibly still depends on the position $x(t)$ of the
particle. For driving field $\varphi=0$ the process is reversible in the
stationary distribution $\sim \exp[-U]$.  The path-space measure
\eqref{action} is formally given, \cite{comment}, by
\[
P_\varphi(\omega) = \exp -\frac 1{4} \int_0^\tau dt \big[\,\big(
\dot{x}(t) + F(x(t)) \big)^2 - \nabla F(x(t))\,\big]
\]
The antisymmetric part under time-reversal in the action $\log P_\varphi$
indeed gives the dissipated power in terms of a stochastic Stratonovich
integral of the ``force'' $G(x(t))$ times the ``velocity'' $\dot{x}(t)$:
$S= \varphi \int G(x(t)) \,\dot{x}(t)\,dt$. On the other hand, the part in
${\mathcal L}_\varphi$ that is symmetric under time-reversal and is
antisymmetric under field-reversal is given by
\begin{equation}\label{rr}
R = \varphi\int_0^\tau \nabla U(x(t))\,G(x(t))\, dt
 +  \frac{\varphi}{2}\int_0^\tau \nabla G(x(t))dt 
\end{equation}
It satisfies the symmetry \eqref{FSS}. When there is a constant external
field $G$, the second term vanishes and the first term picks up the total
force exerted on the particle on its trajectory $x(t), t\in [0,\tau]$. The
same can be repeated when $x(t)=\rho(r,t)$ is a field on some bounded
domain, $r\in V$, that is subject to dissipation and forcing at its `left'
and `right' boundary. The expression of $R$ is likewise of the form$
\int_{\partial_{\ell} V} dr dt U'(\rho(r,t)) - \int_{\partial_r V} dr dt
U'(\rho(r,t))$ where the spatial integration is over the respective
boundaries $\partial_{\ell} V,\partial_{r} V$ of $V$; the integrand
$U'(\rho(r,t))$ could for example indicate the local chemical potential as
defined in the system. In that sense the external force couples with the
local density of the particles. In the case of the previous example (heat
conduction) one has the external force coupling with the local kinetic
energy.

\indent {\bf Conclusions.} For an irreversible thermodynamics as a linear
response theory for time-antisymmetric observables like fluxes or
currents, mostly only the entropy production and its fluctuations are
needed.  Beyond linear order, nonequilibrium physics must come to terms
also with the fluctuations of time-symmetric observables. The symmetry
that we know is present in the fluctuations of the dissipation function is
also found to be valid for a time-symmetric observable ($R$).  That
enables to continue the perturbation expansion around equilibrium beyond
linear order in the driving fields.  As a consequence a Green-Kubo type
relation and corresponding Onsager reciprocity are obtained for
time-symmetric functions ($V$). While some examples were shown to give
specific physical meanings to our $R$ and $V$, still, nonequilibrium
effects beyond linear order reside, at least for theory, in a vast terra
incognita and it will be mainly through experimental work that the
usefulness of our new and  general relations will be checked.

{\it It is a pleasure to
thank Marco Baiesi and Karel  Neto\v cn\'y for valuable
discussions.}

\end{document}